\documentstyle[preprint,prb,aps,epsf]{revtex}

\begin{document}

\tighten

\title{Entangled states in a Josephson charge qubit coupled to a\\
su\-per\-con\-duc\-ting re\-sonator}

\author{O. Buisson$^{1}$ and F.W.J. Hekking$^{2}$}

\address{$^{1}$Centre de Recherches sur les Tr\`es Basses Temp\'eratures,
laboratoire associ\'e \`a l'Universit\'e Joseph Fourier, C.N.R.S., BP 166,
38042 Grenoble-cedex 9, France.\\
$^{2}$ Laboratoire de Physique et Mod\'elisation des Milieux Condens\'es
\& Universit\'e Joseph Fourier, C.N.R.S., BP 166,
38042 Grenoble-cedex 9, France.}
\maketitle

\begin{abstract}
We study the dynamics of a quantum superconducting circuit which
is the analogue of an atom in a high-Q cavity. The circuit
consists of a Josephson charge qubit coupled to a superconducting resonator.
The charge qubit can be treated as a two level
quantum system whose energy separation is split
by the Josephson energy $E_j$. The superconducting resonator in our proposal
is the analogue of a photon box and is described by a quantum harmonic
oscillator with characteristic frequency $\omega_r$. The coupling between the
charge qubit and the resonator is realized by a coupling capacitance $C_c$.
We have studied the eigenstates as well as the dynamics of the quantum
circuit. Interesting phenomena occur when the Josephson energy equals the
oscillator frequency, $E_j=\hbar\omega_r$. Then the quantum circuit is
described by
entangled states. We have deduced the time evolution of these states
in the limit of weak coupling between the charge qubit and the resonator.
We found Rabi oscillations of the excited charge qubit eigenstate.
This effect is explained by the spontaneous emission and re-absorption of a
single photon in the superconducting resonator.
\end{abstract}

\section{Introduction}

Recently, a substantial interest in the theory of quantum
information and computing~\cite{Steane98} has revived the physical research
on quantum systems. The elementary unit of quantum
information is a two-state system, usually referred to as a quantum bit
(qubit).
Basic operations are realized by preparation and manipulation of,
as well as a measurement on,
entangled states in systems which consist of several coupled qubits.
However, the fabrication of physical systems which would enable
the actual implementation of quantum algorithms is far from being
realized in the near future and a substantial amount of fundamental research
is still needed.

During the past five years, great progress has been made in the manipulation
of entangled states in systems consisting of up to four qubits
based on ion traps~\cite{Sackett00} and atoms in a high-Q
cavity~\cite{Haroche98}.
These two experiments demonstrate clearly and unambiguously the possibility
to coherently control the entangled states of a limited number of qubits,
as well as to perform a quantum measurement
on them. In spite of this success,
it seems quite difficult to realize circuits consisting of the large 
number of ion
traps or atoms in a cavity necessary for quantum computation.

It has been suggested that
small solid state devices fabricated using
nanolithography technologies are promising for quantum circuit
integration.
However, the coherent manipulation of entangled states
as well as the realization of quantum measurements remain fundamental issues
to be investigated. One of the main challenges is to gain control over
all possible sources of decoherence.
At present the best candidates for the implementation of quantum gates based
on solid state devices are circuits using small Josephson junctions.
In the superconducting state, such circuits contain less intrinsic sources
of decoherence.
Indeed it has been experimentally demonstrated
that a single Cooper pair box is a macroscopic two level system which can be
coherently controlled~\cite{Bouchiat98,Nakamura97,Nakamura99}.
At about the same time, theoretical works have proposed the
use of the Cooper pair box as a qubit (the so-called Josephson charge qubit)
in the context of quantum computers.
In particular, systems consisting of several charge qubits with controlled
couplings have been discussed, the quantum
measurement problem has been addressed and the decoherence time has been
estimated~\cite{Shnirman97,Maklin99}. More recently,
qubits based on superconducting loops containing small Josephson junctions
have been proposed
(Josephson phase qubits)~\cite{Mooij99,Feigel'man00}
  and
are currently studied~\cite{Vanderwal99,Friedman00}.
But up to now, the existence of entangled states, which are at the heart of
quantum information processing, has been demonstrated neither for
charge qubits nor for phase qubits.

In
this article we propose to study one of the simplest Josephson circuits in
which entangled states can be realized. It consists of a charge qubit coupled
to a superconducting resonator and can be described theoretically by a two
level
system coupled to a quantum harmonic oscillator.
After a description of this quantum circuit in the next section, the
Hamiltonian describing
it will be derived in Sec.~\ref{hamiltonian}. In Sec.~\ref{dynamics},
the time evolution of the eigenstates is obtained and we demonstrate
the existance of entanglement. In the last Section, we discuss the dynamics
of the quantum circuit for typical experimental values
of the system parameters.

\section{Quantum Circuit}
\label{circuit}

\begin{figure}[ht]
\centerline{\epsfxsize=9cm\epsfbox{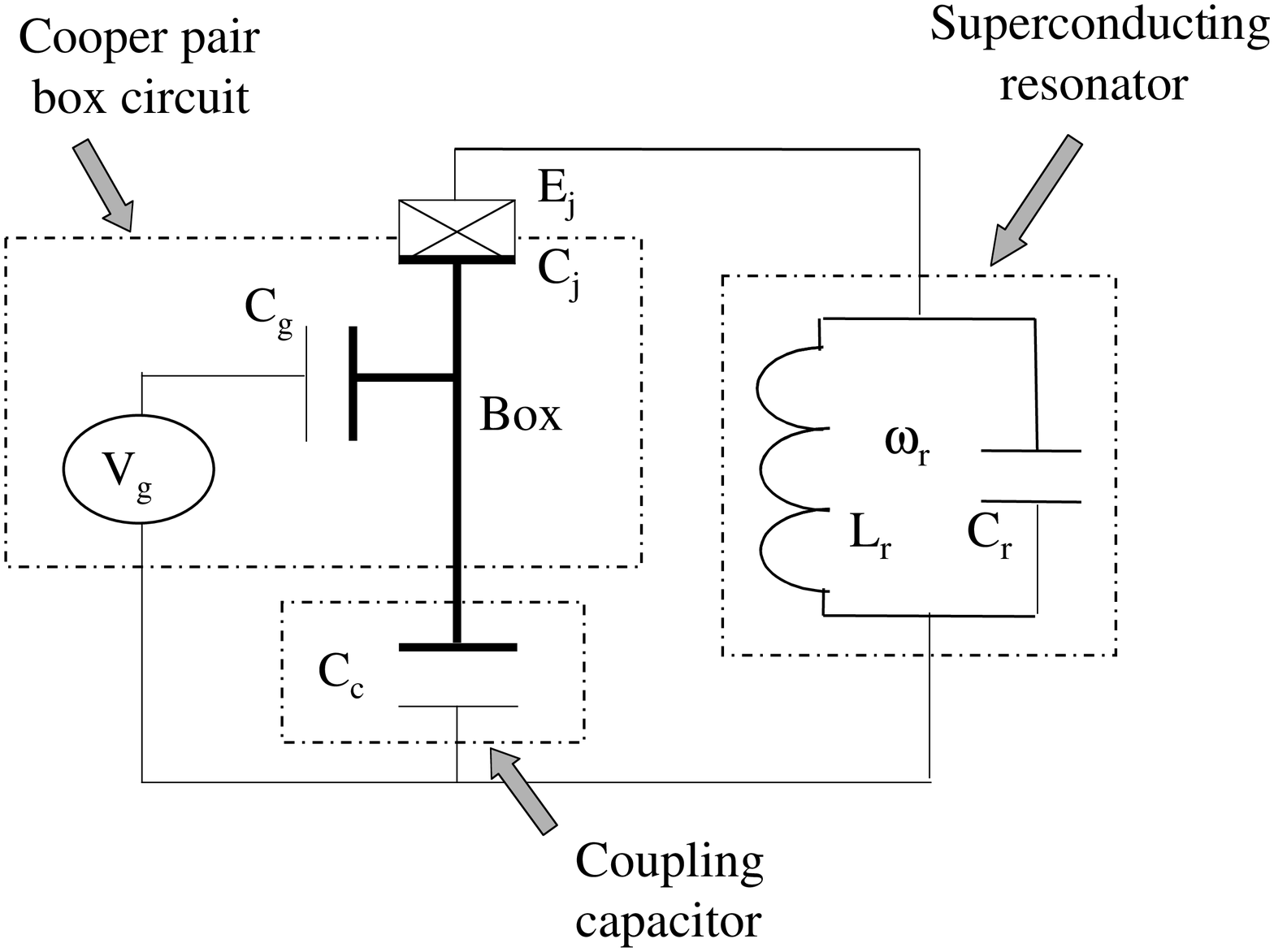}}
\caption{The quantum circuit.}
\label{device}
\end{figure}

The Josephson circuit we study hereafter is depicted in
Fig.~\ref{device}. It consists of three different elementary
circuits: a Cooper pair box, an LC-resonator and a
"coupling" capacitor.

For small enough junction capacitance $C_j$, gate capacitance
$C_{g}$, and coupling capacitance $C_c$, the charging energy
of the box is large
compared to thermal
fluctuations and the excess charge
of the box is quantized.
On the other hand, we assume the charging energy to be smaller than
the superconducting gap $\Delta$, such that no quasiparticles are present in
the box. Thus the excess charge is entirely due to the presence of Cooper
pairs and charge quantization occurs in units of $2e$.
The gate voltage
$V_{g}$ is used as an external control
parameter. When the gate
charge $N_{g} = - C_{g}V_{g}/e$ is equal to unity,
the Cooper pair box can be viewed
as a macroscopic two-level quantum system whose energy separation is split
by the Josephson energy $E_j$. The two eigenstates $|-\rangle$ and $|+\rangle$
correspond to a coherent superposition of the two different charge
states of the box~\cite{Bouchiat98,Nakamura97,Nakamura99}.
When $N_{g} \approx 1$, the Cooper pair box will be referred to
as a Josephson charge qubit.

The LC-resonator system can be described by a
quantum harmonic oscillator whose characteristic frequency is given by
$\omega_r$. This system
is the analogue of a high-Q cavity.

The capacitance $C_{c}$ plays a crucial role in our proposed circuit since it
couples the charge qubit
and the resonator to each other. These two circuits are no longer
independent and
the system must be considered in its totality. Thus the
proposed quantum circuit of Fig.~\ref{device} realizes the simple
situation in which a two level system is coupled to a harmonic
oscillator. In spite of its simplicity, such a system describes a great
variety of interesting situations~\cite{Jaynes63,Brune96,Marquardt00}.

\section{Hamiltonian}
\label{hamiltonian}

The circuit depicted in Fig.~\ref{device} can be characterized mechanically
by two generalized
coordinates, $\phi _{j}$ and $\phi_{r}$.
These coordinates are associated with the voltage drop
$\delta V_{j}$ over the junction and $\delta V_{r}$
over the resonator, respectively, according to the Josephson
relation $\phi _{i} = 2e \delta V_{i} t/\hbar$ ($i=j,r$).
We seek the Lagrangian ${\cal L}(\phi_{j}, \phi_{r}, \dot{\phi}_{j},
\dot{\phi}_r) = T-V$ describing the dynamics of these variables.
The
potential energy $V$ is a function of the coordinates only,
\begin{equation}
V(\phi_{j}, {\phi_{r}})
=
-E_{j} \cos \phi_{j} + \frac{E_{r}}{2} \phi_{r}^{2} ,
\end{equation}
where
$E_{r} = (1/L_{r})(\hbar/2e)^{2}$ is the energy associated
with the inductance $L_r$ of the resonator.
The kinetic energy $T$ is quadratic in the
velocities $\dot{\phi}_{j}$ and $\dot{\phi}_{r}$. It is just
the free electrostatic energy stored in the capacitators present in the
circuit.
This free energy can be written
as
\begin{equation}
T
=
\frac{1}{2}
\left[
C_{\Sigma j} (\hbar \dot{\phi}_{j}/2e)^{2}
+
C_{\Sigma r} (\hbar \dot{\phi}_{r}/2e)^{2}
+
2C_{\Sigma c}(\hbar/2e)^{2} \dot{\phi}_{j} \dot{\phi}_{r}
\right] .
\end{equation}
Here we introduced the capacitances $C_{\Sigma c} = C_{c} + C_g$,
$C_{\Sigma j} = C_{j} + C_{\Sigma c} $, $C_{\Sigma r} = C_{r} +C_{\Sigma c}$.
Note that the
Lagrangian contains an interaction between the resonator and the junction:
${\cal L}_{int} = C_{\Sigma c} (\hbar/2e)^{2} \dot{\phi}_{j} \dot{\phi}_{r}$.
The effective coupling between these two parts of the circuit is determined by
the sum of the gate capacitance and the coupling capacitance.
We also note that the equations of motion $d (\partial {\cal L}/\partial
\dot{\phi}_{i})/dt +
\partial {\cal L}/\partial \phi_{i} = 0$ express current conservation in
the circuit.

Through
the Josephson junction, Cooper pairs can tunnel from or onto the island.
The number of excess Cooper pairs on the island, $n$, depends on the gate
voltage.
Charge neutrality leads us to relation
\begin{equation}
2ne
=
C_{\Sigma j} (\hbar \dot{\phi}_{j}/2e) + C_{\Sigma c} (\hbar
\dot{\phi}_{r}/2e) +  N_g e.
\end{equation}
It is always possible to add a term to the Lagrangian which is a total
time derivative. Let us add the
term $ \hbar \dot{\phi}_{j} N_g /2$. As a result, the momenta
conjugate to $\phi _{j}$ and $\phi _{r}$ are
\begin{eqnarray}
p_j &=& \partial {\cal L}/\partial \dot{\phi}_{j} = (\hbar /2e)
[C_{\Sigma j} (\hbar \dot{\phi}_{j}/2e) + C_{\Sigma c} (\hbar
\dot{\phi}_{r}/2e ) + N_g e] = \hbar n ,\nonumber \\
p_{r} &=& \partial {\cal L}/\partial \dot{\phi}_{r} =
C_{\Sigma r} (\hbar/2e)^{2} \dot{\phi}_{r}
+ C_{\Sigma c} (\hbar/2e)^{2} \dot{\phi}_{j} . \nonumber
\end{eqnarray}
Note in particular that the momentum $p_{j}$ is proportional to $n$.

The Hamiltonian is obtained with the help of the Legendre transform
$H = p_{j} \dot{\phi} _{j} + p_r \dot{\phi}_{r} - { \cal L}$. We find
$H= H_{j} + H_{r} + H_{c}$, where
\begin{eqnarray}
H_{j}
&=&
E_{C,j} (2n - N_{g})^{2} - E_{j} \cos \phi_{j}, \\
H_{r}
&=&
E_{C,r} (2p_{r}/\hbar) ^{2} + E_{r} \phi_{r}^{2} /2, \\
H_{c}
&=&
- E_{C,c} (2 n - N_{g}) (2 p_{r}/\hbar) .
\end{eqnarray}
The charging energies $E_{C,i}$ ($i=j,r,c$) appearing here are given by
$E_{C,i} = e^2/2C_{i, \mathrm{eff}}$, with
$ C_{j,\mathrm{eff}} = C_{j} +
(1/C_{\Sigma c} +1/C_{r})^{-1}$, $C_{r,\mathrm{eff}} = C_{r} +
(1/C_{\Sigma c} +1/C_{j})^{-1}$, and
$ C_{c,\mathrm{eff}}
= [C_{\Sigma c} + (1/C_{j} +1/C_{r})^{-1}][(C_j +C_r)/2C_{\Sigma c}]$.
The Hamiltonian equations of motion, $\dot{p}_{i} = - \partial
H/\partial \phi _{i}$, $\dot{\phi}_{i} = \partial H/\partial p_{i}$
lead us again to current conservation.

In order to obtain the quantum
mechanical Hamiltonian $\hat{H}$, we replace $p_{i}, \phi _{i}$ by the
corresponding operators, with $[p_{k}, \phi_{m}] = (\hbar /i) \delta
_{k,m}$. In particular, in $\phi -$representation we have $p_{k} =
(\hbar/i) \partial /\partial \phi _{k}$. Note also that
$[n,\phi_{j}] = -i$ and $n = -i \partial /\partial \phi _{j}$.
Below we discuss the various contributions to $\hat{H}$ in some detail.

{\em Josephson junction.}
The commutation relation $[n,\phi_{j}] = -i$ implies $[n,e^{i\phi_{j}}]
= e^{i\phi_{j}}$. Using the basis states
$|n\rangle$, where $n$ corresponds to the number of excess Cooper pairs
on the island, we thus have $e^{i\phi_{j}} |n\rangle = |n+1\rangle$.
Similarly, $e^{-i\phi_{j}} |n\rangle = |n-1\rangle$. Therefore we can
write $\hat{H}_{j}$ as
\begin{eqnarray}
\hat{H}_{j}
&=&
E_{C,j} \sum \limits_{n} (2n - N_{g})^{2} |n\rangle \langle n| \nonumber \\
&&-
\frac{E_{j}}{2}
\sum \limits_{n} \left(|n+1\rangle \langle n| + |n-1 \rangle \langle
n| \right).
\label{Hj}
\end{eqnarray}

If the gate-voltage is such that $N_{g} \simeq 1$, the states with
$n=0$ and $n=1$ are almost degenerate. At low temperatures, the
Hamiltonian $\hat{H}_{j}$ involves only these two states, and thus can be
presented as a matrix
\begin{equation}
\hat{H}_{j}
\simeq
\left(
\begin{array}{cc}
E_{C,j}N_{g}^{2} & -E_{j}/2 \\
-E_{j}/2   &E_{C,j} (2-N_{g})^{2}
\end{array}
\right).
\end{equation}
This matrix can be diagonalized. The eigenvalues are
\begin{equation}
E_{\mp}
=
E_{C,j}[1+(\delta N_{g})^{2} ]
\mp
\frac{1}{2} \sqrt{(\delta E_{g})^{2} + E_{j}^{2}},
\end{equation}
where $\delta N_{g} = N_{g} -1$ and $\delta E_{g} = -4E_{C,j} \delta
N_{g}$. The corresponding eigenstates are
\begin{eqnarray}
|-\rangle = \alpha |0\rangle + \beta |1\rangle \\
|+\rangle = \beta |0\rangle - \alpha |1\rangle
\end{eqnarray}
where
$\alpha ^{2} = 1- \beta ^{2} = [1+ \delta E_{g}/\sqrt{
(\delta E_{g})^{2}+ E_{j}^{2}}]/2$.

{\em Resonator.} Since the LC-circuit constitutes just
a harmonic oscillator with a characteristic frequency $\omega _{r}
= \sqrt{1/L_rC_{r,\mathrm{eff}}}$, the Hamiltonian
$\hat{H}_{r}$ can be written in the standard way
\begin{equation}
\hat{H}_{r}
=
\hbar \omega _{r} (a^{\dagger} a +1/2) ,
\label{Hr}
\end{equation}
where
\begin{eqnarray}
\phi _{r} &=& 2 \sqrt{\frac{E_{C,r}}{\hbar \omega _{r}}} (a^{\dagger} +
a), \\
p_{r} &=& \frac{i\hbar}{4}\sqrt{\frac{\hbar \omega _{r}}{E_{C,r}}}
\left(a^{\dagger} - a \right) .
\end{eqnarray}

{\em Coupling term.} The coupling term can also be written using
the operators $a, a^{\dagger}$ :
\begin{equation}
\hat{H}_{c}
=
- i
\frac{E_{C,c}}{2}\sqrt{\frac{\hbar \omega _{r}}{E_{C,r}}}  (2 n - N_{g})
\left(a^{\dagger} - a \right) .
\label{Hc}
\end{equation}
Note that the characteristic coupling energy is
$E_c =  \sqrt{\hbar \omega _{r}/E_{C,r}} E_{C,c}/2$.

A general analysis of the Hamiltonian $\hat{H}$ is beyond the scope of
the present paper and will be presented elsewhere~\cite{Ithier00}.
In the next section
we will discuss an explicit matrix form of $\hat{H}$,
which can be obtained under certain simplifying conditions which are
nevertheless experimentally relevant.

\section{Eigenstates and Entanglement}
\label{dynamics}

Throughout this section we will work in the zero-temperature limit.
We are interested in the situation $N_{g} \simeq 1$,
such that we have to consider the charge qubit states
$|-\rangle$ and $|+\rangle$ only. Furthermore,
as far as the resonator is concerned,
we will consider $\hbar \omega _r = E_j$ and work only with the ground state
$|0\rangle$ and the first
excited state $|1 \rangle$. In the limit of weak coupling,
$E_c \ll \hbar \omega_r$,
the coupled system can be
characterized by the four basis states
$| -, 0\rangle, | -, 1\rangle, | +, 0\rangle, | +, 1\rangle$. The Hamiltonian
matrix for this low-energy subspace reads
\begin{equation}
\hat{H}
=
\left(
\begin{array}{cccc}
  E_0& iE_\beta &0 &-2i \alpha
\beta E_{c} \\
-iE_\beta&E_1
& 2i \alpha \beta E_{c} &0 \\
0 &-2i \alpha \beta E_{c} &E_2
& iE_\alpha\\
2i \alpha \beta E_{c} &0 &-iE_\alpha
&E_3
\end{array}
\right) ,
\label{Hrel}
\end{equation}
which is a hermitian matrix describing the two-level system coupled
to the lowest states of the resonator. Here,
$E_0 = E_{-} + \hbar \omega _{r}/2$,
$E_1 = E_{-} + 3\hbar \omega _{r}/2$,
$E_2 = E_{+} + \hbar \omega _{r}/2$,
$E_3 = E_{+} + 3\hbar \omega _{r}/2$,
$E_\alpha = E_{c}(2\alpha ^{2} - N_{g})$, and
$E_\beta =E_{c}(2\beta ^{2} - N_{g})$.

Suppose that the system has been prepared in the state
$|\psi (t=0)\rangle = |+,0\rangle$ at time
$t=0$. This situation can be achieved by a suitable manipulation of the
gate voltage $V_g$ at times prior to $t=0$~\cite{Nakamura99,Ithier00}.
At times $t>0$, the time
evolution of $|\psi (t)\rangle$ describing the system is governed by the
Hamiltonian~(\ref{Hrel}). We keep $V_g$ fixed such that $N_g = 1$ at $t>0$.
  Thus we have
$\alpha ^2 = \beta ^2 = 1/2$, and hence
$E_\alpha = E_\beta = 0$. Moreover, as $\hbar \omega _r = E_j$,
we have $E_1 = E_2 = E_{C,j}+E_j \equiv \bar{E}$: without
coupling, the state $|+,0\rangle$ would be degenerate with
the state $|-,1\rangle$.
Thus the Hamiltonian takes the simple form
\begin{equation}
\hat{H}
=
\left(
\begin{array}{cccc}
  E_0& 0 &0 &-iE_{c} \\
0&\bar{E}
& i E_{c} &0 \\
0 &-i  E_{c} &\bar{E}
& 0\\
i  E_{c} &0 &0
&E_3
\label{Hsimp}
\end{array}
\right) .
\end{equation}
Note in particular that the state $|+,0\rangle$
couples to the state$|-,1\rangle$;
as a result, the degeneracy between them is lifted
and the states become entangled.
The precise form of the entanglement is governed
by the central $2 \times 2$ block of the matrix~(\ref{Hsimp}).
The eigenstates of this block are
\begin{eqnarray}
|\chi_1\rangle &=& [|-,1\rangle +i |+,0\rangle]/\sqrt{2}, \nonumber \\
|\chi_2\rangle &=& [|-,1\rangle -i |+,0\rangle]/\sqrt{2}, \nonumber
\end{eqnarray}
corresponding to the eigen energies $\bar{E} - E_c$ and
$\bar{E} + E_c$, respectively.
These two excited eigenstates thus correspond to a maximum
entanglement of charge qubit and resonator states,
induced by the capacitive coupling between them.

The time evolution of $|\psi (t) \rangle$ is
given by
\begin{equation}
|\psi (t) \rangle
=
\frac{1}{\sqrt{2}i}
\left[
e^{-i(\bar{E} -E_c)t/\hbar} |\chi_1\rangle
-
e^{-i(\bar{E} + E_c)t/\hbar} |\chi _2\rangle
\right] .
\end{equation}
We deduce that the state $|\psi (t) \rangle$ oscillates coherently between
$|-,1\rangle$ and $|+,0\rangle$. In fact, these so-called
quantum Rabi oscillations can be interpreted as the spontaneous
emission and re-ab\-sorp\-tion of one excitation quantum by the resonator.
An interesting quantity is the probability
$P_1 (t)$ to find the harmonic
oscillator in the state $|1\rangle$ after a certain time $t$.
This probability shows Rabi oscillations as a function of $t$
with frequency $2E_c /\hbar$,
\begin{equation}
P_1 (t)
= |\langle 1,-|\psi(t)\rangle|^2
= \frac{1}{2} [1 - \cos (2E_c t/\hbar)] .
\label{P1}
\end{equation}
Since these Rabi oscillations are characteristic for the
entanglement realized in the system, their measurement would
provide direct evidence of the presence of the entangled states
$|\chi_1\rangle$ and $|\chi_2\rangle$. We will discuss the feasibility
of such a measurement in the next Section.

\section{Discussion}
\label{discussion}

For the numerical estimates presented below we will consider 
parameters related to a typical aluminium
superconducting circuit~\cite{Bouchiat98,Nakamura99}.
For the Josephson charge qubit we have chosen the following characteristics:
$E_j = 26.1 \mu$eV, $E_{C,j} = 70 \mu$eV, $\Delta = 240 \mu$eV.
As for the resonator, we take $L_r = 90$pH and $E_{C,r} = 12$neV,
as a result $\hbar \omega _r = 26.1 \mu$eV. Finally, the coupling capacitance
is chosen to be of the same order of $C_j$, $C_c = 0.5$fF, yielding
$E_c = 256$neV. Note that the coupling energy is indeed much smaller than the
Josephson energy, which in turn is equal to the excitation energy
of the resonator.

\begin{figure}[ht]
\centerline{\epsfxsize=7cm\epsfbox{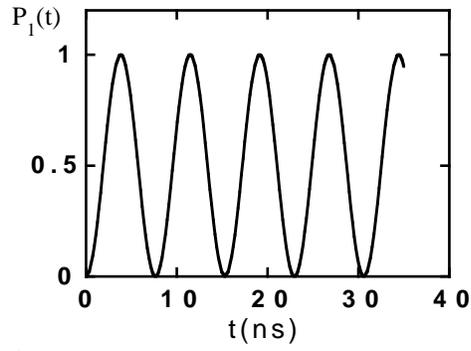}}
\caption{Probability $P_1(t)$ to find the system, prepared in the state
$|+,0\rangle$ at $t=0$, in the state $|-,1\rangle$
after a time $t$.}
\label{rabi}
\end{figure}

Using the above paramaters, we have plotted $P_1(t)$, Eq.~(\ref{P1}),
as a function of time in Fig.~\ref{rabi}. We clearly see the Rabi oscillations
with perdiodicity $T_{\rm Rabi} \approx $8 ns.

In order to be able to observe these oscillations, we need to satisfy
various conditions. First of all, in order to avoid the presence
of quasiparticles, the temperature must be much lower than the
gap $\Delta$.
Secondly, it is necessary to have a
decoherence time which is longer than $T_{\rm Rabi}$.
In our system, the decoherence time will be the shorter
of the lifetime $\tau _r$ of the excitated state of the resonator and
the decoherence time $\tau _{\rm qubit}$ of the charge qubit.
A Q-factor of about 500 is quite realistic for a superconducting
LC resonator. This yields a lifetime $\tau _r >$ 10 ns.
As for the qubit, the experiment by Nakamura~\cite{Nakamura99} has
indicated that $\tau _{\rm qubit} >$ 2ns. This lower bound
is essentially the decoherence time
associated with the coupling to the measuring probe.
In principle, this time can be improved upon by modifications of 
the experimental set-up.  
Theoretical estimates~\cite{Maklin99} show that a time
$\tau _{\rm qubit} \sim$ 100 ns should be feasable.
Finally, a measurement of the number of excitations
should be performed on the resonator. This can be done, {\em e.g.},
along the lines of Ref.~\cite{Devoret85}, where the discrete,
oscillator-like energy levels of an underdamped Josephson junction
were measured.

\acknowledgments{ We thank F. Balestro, M.M. Doria, G.
Ithier, J.E. Mooij and J. M. Raimond for invaluable discussions.}

\end{document}